\documentclass[openacc]{rsproca_new}

\usepackage{bm}

\titlehead{Comment}

\begin{document}

\title{Comment on `On computing quantum waves exactly from classical action'}

\author{G\'abor Vattay$^{1}$}

\address{$^{1}$Department of Physics, E\"otv\"os Lor\'and University,
H-1053 Budapest, Egyetem t\'er 1--3, Hungary\\
ORCID: 0000-0002-0919-9429}

\subject{quantum physics, mathematical physics}


\corres{G\'abor Vattay\\
\email{gabor.vattay@ttk.elte.hu}}

\begin{abstract}
A recent article by Lohmiller \& Slotine (Proc.\ R.\ Soc.\ A \textbf{482}: 20250413)
claims that the Schr\"odinger equation can be solved exactly using only classical least action
and classical fluid density, asserting that this formulation avoids semiclassical approximations.
We show that their mathematical derivation contains a foundational error. By neglecting
the spatial derivatives of the probability density amplitude, the authors inadvertently
omit the quantum potential---the term originally identified by Madelung and later
emphasised by Bohm. Consequently, their proposed equivalence is not exact but rather
constitutes the standard semiclassical approximation. We further demonstrate that each
of the paper's illustrative examples either belongs to a class where the quantum potential
vanishes identically due to the geometry of the problem, or recovers the correct quantum
result by importing quantum eigenfunctions through the initial conditions, thereby concealing
the error.
\end{abstract}

\begin{fmtext}

\section{Introduction}\label{sec:intro}

In their publication, Lohmiller \& Slotine~[1] propose that the quantum wave function
\begin{equation}\label{eq:ansatz}
\psi_j = \sqrt{\rho_j}\, e^{i\varphi_j/\hbar},
\end{equation}
constructed from the classical action~$\varphi_j$ and the classical density~$\rho_j$ evaluated along
a classical path~$j$, is an exact solution to the Schr\"odinger equation. The central claim
rests on their lemma~3.1, where substituting~(\ref{eq:ansatz}) into the Schr\"odinger equation
is asserted to yield exactly the Hamilton--Jacobi partial differential equation, with no
remainder. Their proof states: ``The density~(2.6) is only defined along each individual
path~$x(t)$, so that its total differential has no variation with respect to~$x$.''

\end{fmtext}
\maketitle

This statement conflates the total derivative along a trajectory with the spatial partial
derivatives required by the kinetic energy operator in the Schr\"odinger equation. The
Schr\"odinger equation is a spatial partial differential equation; the Laplacian~$\nabla^2$
must act on the full spatial dependence of~$\psi_j$, including~$\sqrt{\rho_j}$. By treating
the density as having no spatial variation, the authors implicitly set
$\nabla\sqrt{\rho_j} = 0$, thereby eliminating the quantum potential. This omission is
precisely the defining feature of the semiclassical (WKB) approximation~[2--5],
not an exact result.

\section{Mathematical correction}\label{sec:math}

To verify the claim, one must substitute the ansatz~(\ref{eq:ansatz}) into the
time-dependent Schr\"odinger equation. For simplicity, and without loss of generality
regarding the central argument, we consider the case of a scalar potential~$V(x)$ with
unit metric~$M$:
\begin{equation}\label{eq:TDSE}
\left[i\hbar\,\frac{\partial}{\partial t} + \frac{\hbar^2}{2M}\,\nabla^2 - V\right]\psi_j = 0.
\end{equation}
The kinetic energy operator involves the spatial Laplacian~$\nabla^2$. Applying the standard
product and chain rules of multivariable calculus to~$\psi_j$ yields the spatial gradient
\begin{equation}\label{eq:grad}
\nabla\psi_j = \left(\nabla\sqrt{\rho_j} + \frac{i}{\hbar}\,\sqrt{\rho_j}\,\nabla\varphi_j\right)
e^{i\varphi_j/\hbar},
\end{equation}
and subsequently the full Laplacian
\begin{equation}\label{eq:laplacian}
\nabla^2\psi_j = \left[\nabla^2\!\sqrt{\rho_j}
+ \frac{2i}{\hbar}\,\nabla\!\sqrt{\rho_j}\cdot\nabla\varphi_j
+ \frac{i}{\hbar}\,\sqrt{\rho_j}\,\nabla^2\varphi_j
- \frac{1}{\hbar^2}\,\sqrt{\rho_j}\,(\nabla\varphi_j)^2
\right] e^{i\varphi_j/\hbar}.
\end{equation}
When this expression is substituted back into equation~(\ref{eq:TDSE}) and separated into
real and imaginary components, the \emph{imaginary part} yields the continuity equation
\begin{equation}\label{eq:continuity}
\frac{\partial\rho_j}{\partial t} + \nabla\cdot\!\left(\rho_j\,\frac{\nabla\varphi_j}{M}\right) = 0,
\end{equation}
which is indeed satisfied by the classical density. However, the \emph{real part} dictates
the energy conservation of the system:
\begin{equation}\label{eq:QHJ}
\frac{\partial\varphi_j}{\partial t} + \frac{1}{2M}\,(\nabla\varphi_j)^2 + V
- \frac{\hbar^2}{2M}\,\frac{\nabla^2\!\sqrt{\rho_j}}{\sqrt{\rho_j}} = 0.
\end{equation}
The last term in equation~(\ref{eq:QHJ}) is the well-known \emph{quantum potential}
\begin{equation}\label{eq:Qpot}
Q = -\frac{\hbar^2}{2M}\,\frac{\nabla^2\!\sqrt{\rho_j}}{\sqrt{\rho_j}},
\end{equation}
originally identified by Madelung in his hydrodynamic formulation of quantum
mechanics~[6] and later popularised by Bohm~[7]. If, and only if, the quantum potential
vanishes does equation~(\ref{eq:QHJ}) reduce to the classical Hamilton--Jacobi equation.

The derivation presented by Lohmiller \& Slotine~[1] implicitly assumes
$\nabla\sqrt{\rho_j} = 0$, thereby artificially eliminating the quantum potential. This
assumption is equivalent to requiring that the probability density has no spatial gradient
whatsoever---i.e.\ that the particle's distribution is perfectly homogeneous across all of
space. For any physical, localised quantum state such as a Gaussian wave packet, the
density varies spatially, the quantum potential is explicitly non-zero, and the classical
Hamilton--Jacobi equation cannot serve as an exact equation for the phase of the wave
function.

The omission of the quantum potential is not a mechanism for obtaining exact solutions,
but rather the defining characteristic of the semiclassical limit. The relationship between
classical action, fluid density and the quantum wave function is a foundational aspect
of semiclassical quantum mechanics, exhaustively detailed in the established
literature~[2--5].

\section{Why the examples appear to work}\label{sec:examples}

If the derivation is mathematically flawed for a single classical branch, how is it that
the examples in~[1] reproduce the correct, textbook quantum-mechanical results? The
resolution lies in the fact that the illustrative examples fall into two distinct categories:
either the quantum potential vanishes identically due to the geometry of the problem,
or the quantum structure is imported through the initial conditions via an infinite
superposition of classical paths.

\subsection{Trivial density: the quantum potential vanishes}\label{sec:trivial}

In certain physical configurations, the classical probability density is spatially
uniform (or varies as~$1/r$ in three dimensions), so that
$\nabla\sqrt{\rho_j} = 0$ holds exactly. In these special cases the assumption made in~[1]
happens to be correct, and the classical Hamilton--Jacobi equation coincides with
the quantum Hamilton--Jacobi equation~(\ref{eq:QHJ}).

1. \emph{Particle in a box and quantum tunnelling} (examples~3.7 and~3.8 in~[1]).
In these systems, the authors work with plane waves of constant momentum between
the walls or outside the barrier. The classical density~$\rho$ is spatially constant. Since
the derivative of a constant is zero, the quantum potential vanishes identically, and
the mathematical error produces no visible discrepancy.

2. \emph{The double-slit experiment} (example~3.5 in~[1]). The classical density of
particles passing through a slit decreases with distance as
$\rho \propto 1/r^2$ in three dimensions, so that $\sqrt{\rho} \propto 1/r$. From standard
vector calculus, the Laplacian of~$1/r$ vanishes everywhere except at the origin
(the slit itself), where it is singular. In the free space beyond the slits, the quantum
potential is therefore identically zero, and the Fraunhofer diffraction pattern
is recovered correctly.

These examples do not test the general validity of the framework; they merely confirm
it in the subset of problems where the omitted term happens to be zero.

\subsection{Circular reasoning: quantum eigenfunctions in the initial conditions}\label{sec:circular}

For bound states with spatially varying densities---such as the harmonic oscillator
(example~3.9) and the hydrogen atom Coulomb potential (example~3.10)---the quantum
potential is manifestly non-zero, and the error in the single-branch derivation cannot
be circumvented. In these examples, the authors do not rely on a single classical
branch but instead sum over an infinite ensemble of initial conditions using equation~(3.3)
in~[1].

The key observation is that this summation reintroduces the full quantum-mechanical
structure through the back door:

1. \emph{Harmonic oscillator.} The authors decompose the initial classical density
distribution in a Taylor series (their equation~3.11) and identify the expansion
coefficients with Hermite polynomials~$H_k$. However, the Hermite polynomials
multiplied by the Gaussian weight are precisely the energy eigenfunctions of the
quantum harmonic oscillator~[8]. The classical action then provides only the
time-dependent phase rotation~$e^{-iE_k t/\hbar}$ for each eigenfunction. In other words,
the quantum basis functions are not derived from classical mechanics but are
\emph{assumed} in the parameterisation of the initial density distribution.

2. \emph{Hydrogen atom / Coulomb potential.} Similarly, in the Duru--Kleinert
transformed coordinates~[9], the authors expand the initial density over products
of Hermite polynomials (their equation~3.16), which correspond, upon
back-transformation, to the spherical harmonics and radial wave functions of the
hydrogen atom~[10]. Once again, the quantum eigenfunctions are imported through the
choice of basis for the initial conditions.

The logic is thus circular: rather than deriving the quantum wave function from
classical mechanics, the authors decompose the initial classical density in a basis that
already consists of the quantum eigenfunctions, and then use the classical action merely to
propagate the phases. The summation over the infinite classical ensemble masks the
single-branch error precisely because it is mathematically equivalent to expanding the
wave function in its own eigenbasis---a standard quantum-mechanical procedure.

\section{Conclusion}\label{sec:conclusion}

The relationship between classical action and quantum waves is a thoroughly mapped
domain in semiclassical mechanics. Berry \& Mount~[2] provide a comprehensive
treatment of how classical action relates to the quantum wave function through precisely
such approximations. Standard derivations of semiclassical evolution~[3,4] demonstrate that
classical trajectories serve to transport the probability density, but exact wave
propagation requires the full inclusion of spatial dispersion through the quantum
potential. Madelung's fluid dynamics naturally accounts for this inherent
dispersion~[3,6].

The assertion that one can bypass these established limits simply by dropping the
spatial derivatives of the density amplitude introduces confusion into a topic
that is well understood. The framework presented in~[1] represents a standard
semiclassical approximation, not an exact mathematical equivalence. Its examples
either belong to the special class where the semiclassical approximation is exact (trivial
density) or implicitly import the quantum solution through the initial conditions.

\vskip6pt


\ack{%
\textbf{Data accessibility.} This article has no additional data.\\[3pt]
\textbf{Declaration of AI use.} AI-assisted tools were used in the preparation of this manuscript.\\[3pt]
\textbf{Authors' contributions.} G.V.\ conceived the analysis, performed the calculations and wrote the manuscript.\\[3pt]
\textbf{Conflict of interest declaration.} The author declares no competing interests.\\[3pt]
\textbf{Funding.} This research was carried out with the support of the
Ministry of Culture and Innovation, funded by the National Research Development and
Innovation Fund, under project number 2022-2.1.1-NL-2022-00004.}


\end{document}